\documentstyle[graphicx,prl,aps,twocolumn]{revtex}
\begin{document}
\title{ Regularities in football goal distributions}
\author{ L. C. Malacarne $\&$ R. S. Mendes}
\address{Departamento de F\'\i sica, Universidade Estadual de Maring\'a,
Avenida Colombo 5790, 87020-900,
 Maring\'a-PR, Brazil}
\maketitle

\begin{abstract}
Besides of complexities concerning to football championships, it
is identified some regularities in them. These regularities refer
to goal distributions by goal-players and by games. In particular,
the goal distribution by goal-players it well adjusted by the
Zipf-Mandelbrot law, suggesting a conection with an anomalous
decay.
\end{abstract}
\date{\today }

 Regularity in  some complex  systems can sometimes  be
identified and expressed   in terms of simple laws. Typical
examples of such situations are found in a wide range of contexts
as the frequency of words in a long text\cite{Zipf}, the
population distribution in big cities\cite{h1,2,3}, forest
fires\cite{5}, the distribution of species lifetimes for North
American breeding bird populations\cite{h2}, scientific
citations\cite{6,7}, www surfing\cite{8}, ecology\cite{8a}, solar
flares\cite{8b}, economic index\cite{9}, epidemics in isolated
populations\cite{Rhodes}, among others. Here, universal behaviours
in the most popular sport, the football, are discussed. More
precisely, this work focuses on regularities in goal distribution
by goal-players and by games in championships. Furthermore, the
goal distribution by goal-players is connected with an anomalous
decay related to the Zipf-Mandelbrot\cite{Zipf,Mandelbrot} law and
with Tsallis nonextensive statistical
mechanics\cite{Tsallis,Curado,Denisov,Tsallis2}.

In many contexts, it is common that few phenomena with high
intensity arise, and  so do many ones with low intensity. For
instance, a long text generally contains many words that are
employed in few opportunities and  a small number that occurs
largely\cite{Zipf}. The above mentioned systems are good examples
too. In particular, this kind of behaviour usually  occurs in
football championships, because there are many players that make
few goals in contrast with the topscorers.

\begin{figure}
 \centering
 \DeclareGraphicsRule{ps}{eps}{*}{}
 \includegraphics*[width=10cm, height=7cm,trim=2.5cm 2.35cm 0cm 2cm]{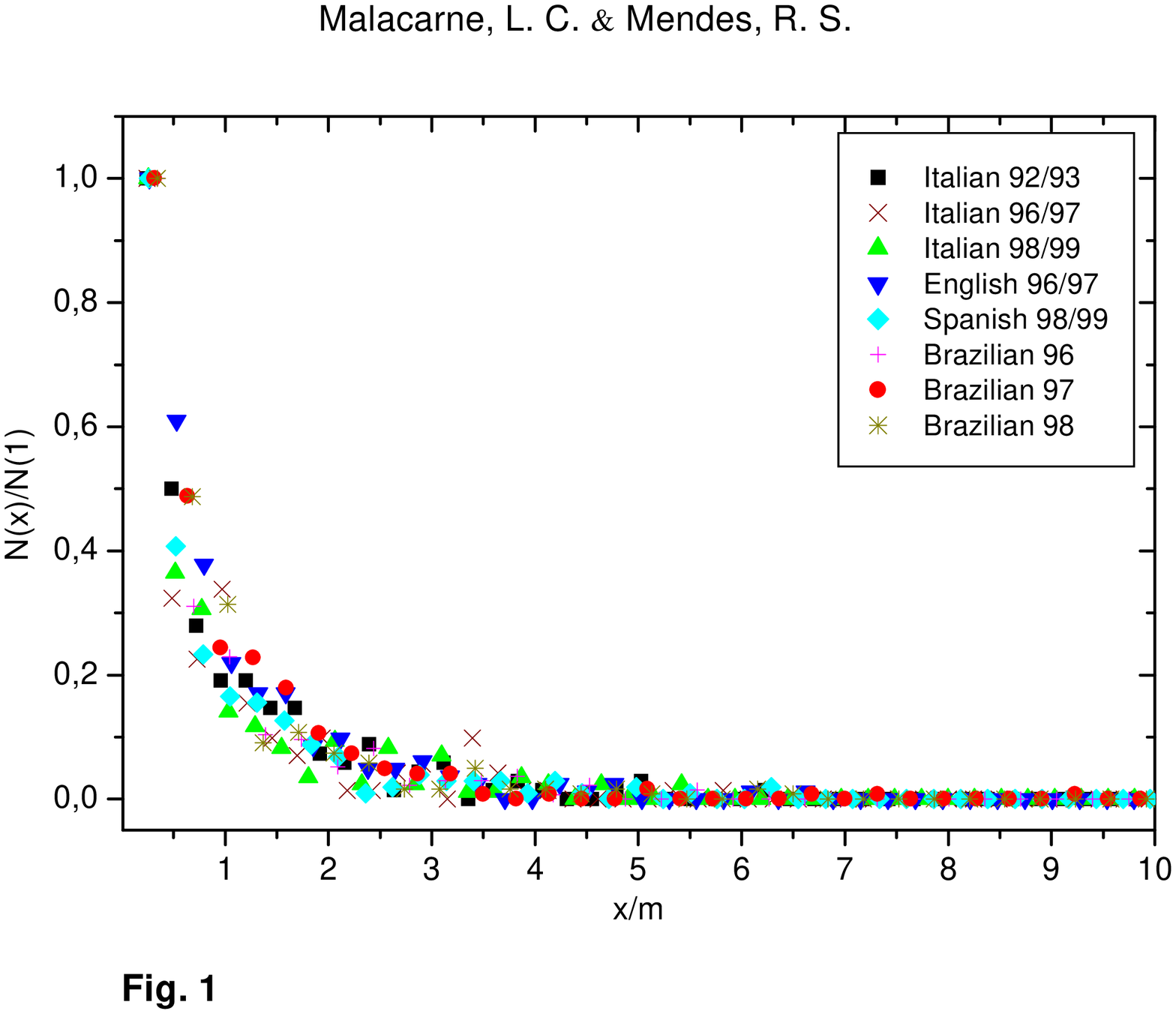}
  \caption{Scaled distribution of goals in main league football
   championships from Italy, England, Spain and Brazil.
  The ordinate $N(x)$ is the number of players
  with $x$ goals divided by the number of players with one goal, $N(1)$,
  and the abscissa is the number of goals divided by the average of goals
  $m$. These scaled data indicate a regularity in the goal
  distribution for goal-players. The European championships
  start in one year and finish in the subsequent year,
  and the  Brazilian championships start and finish in the same
  year.
  }\label{f2}
\end{figure}

A detailed visualization of this behaviour can be well illustrated
by considering some of the most competitive and traditional
championships of the world. Our particular choice of championships
has been done guided by the criterion of easy accessibility of the
corresponding data to anyone\cite{data1,data2}. Therefore, we
consider, here, some of the main league football championships
from Italy, England, Spain and Brazil\cite{obs1}. Each of these
championships has the participation of about twenty teams,
contains around three hundred games, and approximately eight
hundred goals\cite{obs2}. In Fig. \ref{f2} we exhibit data of
these championships. In these graphics, the abscissa presents the
number of goals $x$ divided by an average of goals $m$ (total
number of goals per total number of goal-players), and the
ordinate indicates the quantity $N(x)$ of players with $x$ goals
divided by the quantity of players with one goal, $N(1)$. The
regular shape of the graphics presented in Fig. \ref{f2} suggests
a general law to describe the distribution of goals.

In the study of the majority of the previously cited systems, the
Zipf's law\cite{Zipf}, $N(x)=a/x^b$, arises naturally, at least in
part of the analysis. In the Zipf's law, $a$ and $b$ are constants
and $x$ is the independent variable. In order to give a better
adjustment to a large part of the data, and based on information
theory, Mandelbrot\cite{Mandelbrot} proposed $N(x)=a/(c+x)^b$ as a
generalization of the Zipf's law, with $a$, $b$, and $c$ all being
constants. This Zipf-Mandelbrot's distribution also arises in the
context of a generalized statistical mechanics proposed some years
ago\cite{Tsallis,Curado,Tsallis2}, equivalently rewritten as
\begin{equation}\label{m}
N(x)= N_0 [1-(1-q)\lambda x]^{\frac{1}{1-q}}\; ,
\end{equation}
where $N_0$, $\lambda$, a $q$ are  real parameters. In addition,
this function satisfies an anomalous decay equation,
\begin{equation}\label{1}
\frac{d}{d x} \left(\frac{N(x)}{N_0}\right) =-\lambda
\left(\frac{N(x)}{N_0}\right)^q \; .
\end{equation}
 The parameter $q$ can be considered as a measure of how
anomalous  the decay is. In particular, equation (\ref{m}) is
reduced to the usual exponential decay, $N(x)= N_0 \exp (-\lambda
x)$, in the limit $q\rightarrow 1$.

\begin{figure}
 \centering
 \DeclareGraphicsRule{ps}{eps}{*}{}
 \includegraphics*[width=9.5cm, height=7cm,trim=2cm 1.8cm 0cm 1.5cm]{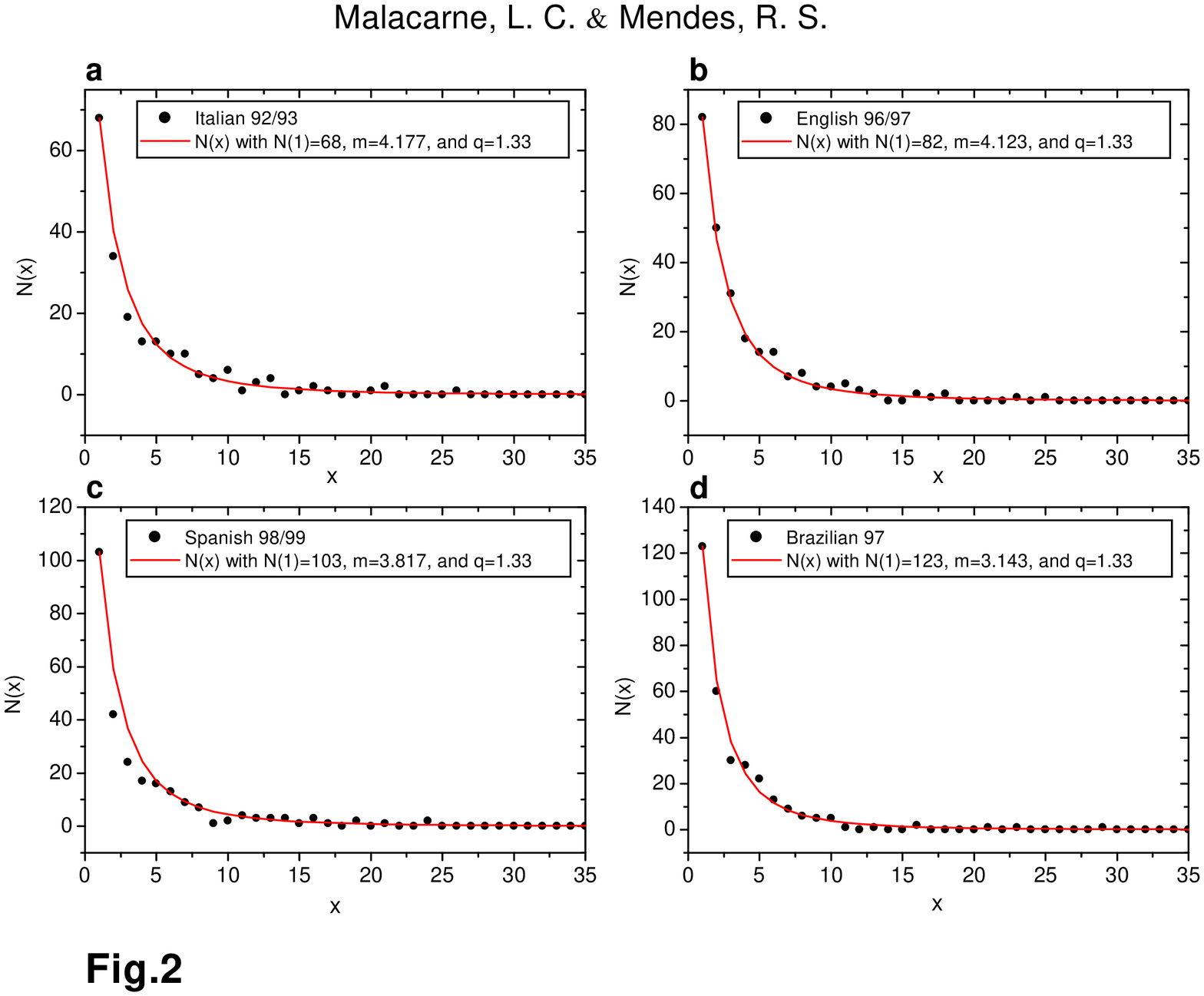}
  \caption{Fit of the goal distribution for goal-players in Italian ({\bf a}),
  English ({\bf b}), Spanish ({\bf c}), and Brazilian ({\bf d}) championships.
  The black circles are the goal data and the solid line is the fitting with the $N(x)$
  distribution being given in equation (\ref{4}) with $q=1.33$. } \label{f3}
\end{figure}

Motivated by these physical connections, we employ the
distribution (\ref{m}) to adjust the goals data. Following the
construction of Fig. \ref{f2}, we use the number of goalplayers
with one goal, $N(1)$, and the average goal number by goalplayer,
$m$, to eliminate $N_0$ and $\lambda$. Furthermore, it is a good
approximation to replace the discrete average with a continuous
one in the present analysis, {\it i.e.},
\begin{equation}\label{3}
  m= \frac{\int_0^{\infty} x N(x)}{\int_0^{\infty}  N(x)}=
  \frac{1}{\lambda (3-2 q)}\;\;\;\;\;\;\; (q<3/2) .
\end{equation}
Thus,  the distribution of goals dictated by equation (\ref{m})
can be rewritten as
\begin{equation}\label{4}
  N(x)=N(1)\frac{ \left[1-\frac{(1-q)}{(3-2q)m} \right]^{\frac{1}{q-1}}}
  {\left[1-\frac{(1-q)}{(3-2q)m}x \right]^{\frac{1}{q-1}}}\; ,
\end{equation}
where $q$ becomes the unique parameter that remains to be
adjusted, since $N(1)$ and $m$ are obtained directly from the
data. Fig. \ref{f3} illustrates applications of equation (\ref{4})
for four championships, indicating therefore the goodness of the
formula (\ref{4}). The same conclusion is obtained in the other
championships showed in Fig. \ref{f2}. Here, $q= 1.33$  was
employed as approximated value, leading to the Zipf-Mandelbrot's
exponent $b \approx 3$. In this way, $q\approx 1.33$ can be
interpreted as the universal parameter for this kind of
championships. Also, it is interesting to remark that $b\approx 3$
occurs in the distribution of scientific citations \cite{6,7}.

\begin{figure}
 \centering
 \DeclareGraphicsRule{ps}{eps}{*}{}
 \includegraphics*[width=10cm, height=7cm,trim=2cm 1cm 0cm 1.5cm]{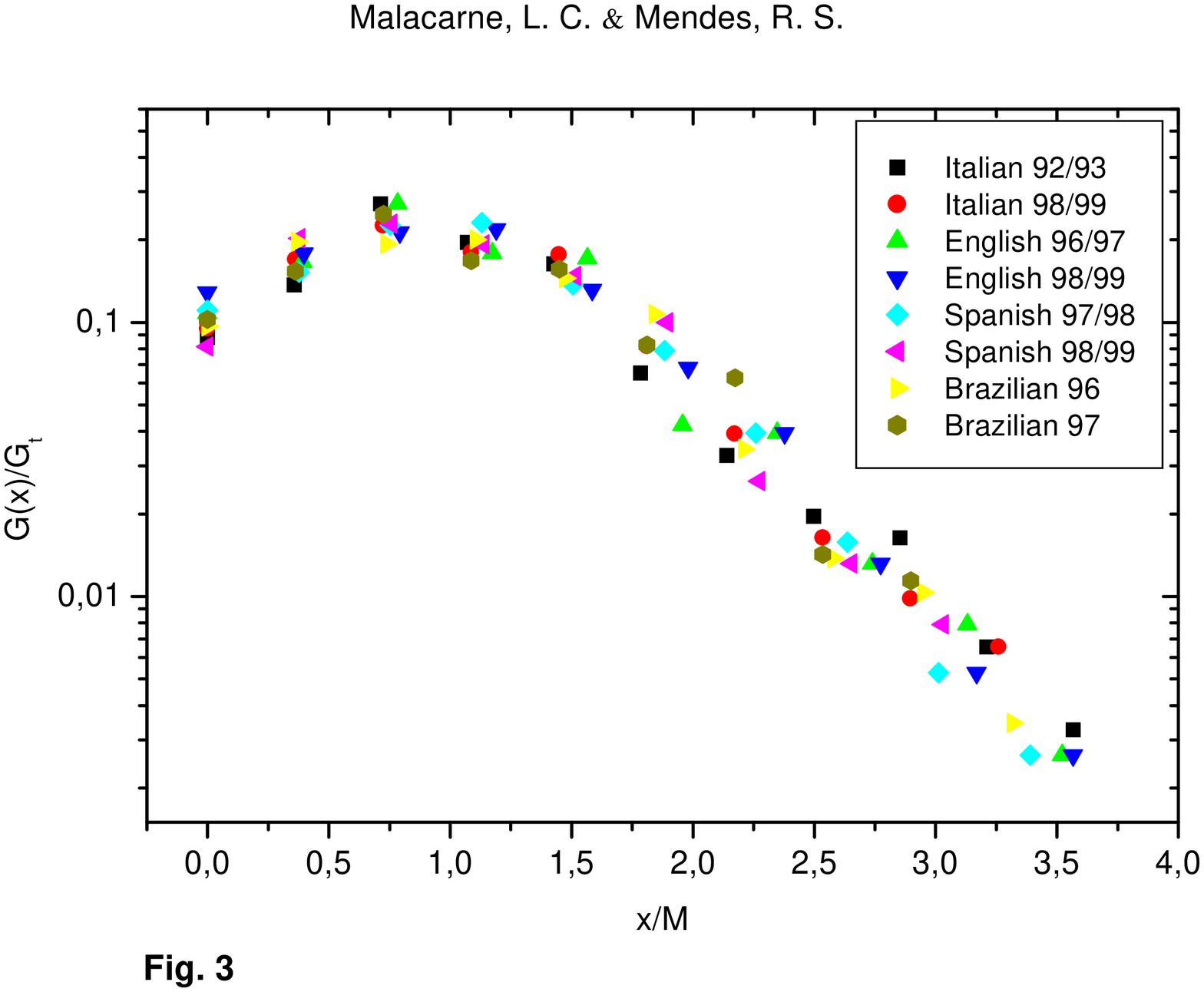}
   \caption{Semi-log graphic of scaled distributions of goals per game
  is illustrated by considering championships from Italy, England,
  Spain and Brazil. For each championship, $x$ is the number
  of goals per game, $M$ is the average of goals per game,
   $G(x)$ is the number of games with $x$ goals, and
   $G_{t}$ is the total number of games. }\label{f4}
\end{figure}

Other kinds of regularities in goal distributions can be
identified, but with different behaviours. This fact can be
verified in the distribution of goals per game. Proceeding in a
similar way as done in Fig. \ref{f2}, it is considered normalized
scale distribution of games and goals. In this case, the abscissa
is the number $x$ of goal divided by $M$, the mean goal per game
of a championship (the number of goals of a championship divided
by the corresponding number of games). In addition, the ordinate
is given by the number of games with $x$ goals of a championship
divided by the number of games of the corresponding championship.
Fig. \ref{f4} contains this kind of graphics illustrating this
regular behaviour by considering, again, the main league football
championships from Italy, England, Spain and Brazil. As one can
see, this figure strongly suggests a regularity in the
distribution of goals for distinct championships.

Besides  the numberless factors, including fluctuation due to
relatively small number of teams and games that are present in a
football championship, regularities arise in the goal
distributions. In particular, the goal distribution for players
that make goals are well adjusted by a Zipf-Mandelbrot law,
suggesting a connection with ubiquitous phenomena such as
anomalous diffusion.


\acknowledgements One of us, Mendes, R. S.,  thanks partial
finantial support by CNPq (Brazilian Agency).



\begin{thebibliography}{10}
\bibitem{obs1}
Since the complete set of data for all championships of these
countries is not easely avaliable, we have used data that are both
rencent and easely accessible.
\bibitem{obs2}
In the European championships, the leagues are organized in such
way that each team plays two games with each of the other teams.
In the Brazilian championships, each team plays only once with
each one of the others. After this, the eight best teams go to the
playoff. Nevertheless, the number of games of the Brazilian
championships in the playoff is much smaller than in the
classification phase, and it doesn't affect the analysis when
compared with European ones.
\end{thebibliography}
\end{document}